# Biophysical characterization of DNA origami nanostructures reveals inaccessibility to intercalation binding sites

Helen L . Miller[1,2], Sonia Contera[3], Adam J.M. Wollman[1,4,5], Adam Hirst[6], Katherine E. Dunn[7,8], Sandra Schröter[6], Deborah O'Connell[6],   Mark C. Leake[1,4]

[1] Department of Physics, University of York, Heslington, York, YO10 5DD, United Kingdom.
[2] Current address: Department of Physics, Clarendon Laboratory, University of Oxford, Oxford, OX1 3PU, United Kingdom.
[3] Department of Physics, Clarendon Laboratory, University of Oxford, Oxford, OX1 3PU, United Kingdom.
[4] Department of Biology, University of York, Heslington, York, YO10 5NG, United Kingdom.
[5] Current address: Biosciences Institute, Newcastle University, NE1 7RU, United Kingdom.
[6]York Plasma Institute, Department of Physics, University of York, Heslington, York, YO10 5DQ, United Kingdom.
[7]Department of Electronics, University of York, Heslington, York, YO10 5DD, United Kingdom.
[8] Current address: School of Engineering, Institute for Bioengineering, University of Edinburgh Faraday Building, King's Buildings, Colin Maclaurin Road, Edinburgh, EH9 3DW, United Kingdom.

E-mail: mark.leake@york.ac.uk

**Abstract**

Intercalation of drug molecules into synthetic DNA nanostructures formed through self-assembled origami has been postulated as a valuable future method for targeted drug delivery. This is due to the excellent biocompatibility of synthetic DNA nanostructures, and high potential for flexible programmability including facile drug release into or near to target cells. Such favourable properties may enable high initial loading and efficient release for a predictable number of drug molecules per nanostructure carrier, important for efficient delivery of safe and effective drug doses to minimise non-specific release away from target cells. However, basic questions remain as to how intercalation-mediated loading depends on the DNA carrier structure. Here we use the interaction of dyes YOYO-1 and acridine orange with a tightly-packed 2D DNA origami tile as a simple model system to investigate intercalation-mediated loading. We employed multiple biophysical techniques including single-molecule fluorescence microscopy, atomic force microscopy, gel electrophoresis and controllable damage using low temperature plasma on synthetic DNA origami samples. Our results indicate that not all potential DNA binding sites are accessible for dye intercalation, which has implications for future DNA nanostructures designed for targeted drug delivery.

## Introduction

Over the past decade important progress has been made in the practical use of DNA nanostructures for targeted therapeutic drug delivery [1–5], in particular for intercalating molecules [6–9], including the anti-cancer drug doxorubicin (DOX). To deliver high doses of a drug it is desirable to have a high initial loading into the DNA carrier nanostructure, high uptake to the desired target area, and a high release rate in the vicinity of a target cell.

DNA nanostructures loaded with the leukaemia drug Daunorubicin [10] have been shown to reduce the viability of target cells more than the same amount of free compound, but questions about the optimum design of the DNA nanostructure carrier for drug loading and cellular uptake remain. Many previous studies have used DNA origami with cage-like structures, but the use of compact structures with intercalated drug molecules could enable higher drug loading, leading to more efficient delivery of the active agent to the target cells.

Well-defined DNA structures instead of linear configurations might be more robust to damage mechanisms encountered in the body. For instance, a recent study showed that tetrahedral DNA nanostructures did not cause significant impairment to native physiology [11], triggering interest in DNA structures with an internal column for drug transportation. Another recent work [12] has shown that large (~MDa) DNA nanoparticles with a high external surface area-to-volume ratio are preferentially taken up in mammalian cell lines relevant to therapeutic drug delivery. Zeng et al. [9] found that rigid 3D DNA origami shapes are more readily taken up by cells, and exhibit sustained drug release, compared to more flexible 2D DNA structures.

Here, we investigate the interactions of intercalating molecules with a dense DNA origami structure using the well-studied YOYO-1 and acridine orange dyes. Previous work with the predominantly bis-intercalating YOYO-1 [13] has focussed on interactions with linear DNA (see for example [14–17]), whilst acridine orange is often used to determine whether DNA is single or double-stranded [18]. Acridine orange binds to double stranded DNA (dsDNA) via an intercalative mode [19,20] and to single-stranded DNA (ssDNA) via electrostatic binding to phosphates [21], producing green or red fluorescence respectively (DNA binding modes are schematically illustrated in figure 1a).

The DNA origami tile used in this work is adapted from the work of Rothemund [22]; the original design uses the viral ssDNA of m13mp18ss as a template and 216 staple strands of 32 nt in length to fold it into a rectangle approximately 70 x 100 nm in lateral dimensions with a central seam, as can be seen in the AFM images in figure 1d, e. For this work the DNA origami tile has been modified to include four strands with a 5' 4-T linker and biotin for surface attachment which all protrude on the same side (figure 1b, and supplementary information) and the addition of 4-T loops to 24 of the staples to inhibit edge stacking, as has previously been described for other DNA origami shapes [22]). Three of the original staples were replaced with 4 alternative staples, one of which is extended (supplementary information), for the purposes of an unrelated experiment that will not be discussed here.

We have used gel electrophoresis, fluorescence microscopy and atomic force microscopy (AFM) to investigate the loading of intercalating molecules into DNA origami with a compact and highly linked structure. To controllably damage the DNA tiles we used low temperature plasma (LTP, figure 1c), to investigate how damage to the DNA structure changes the available intercalation sites. LTP, a partially ionized gas and electrically neutral state of matter, is known to induce DNA damage via single- and double-strand breaks [23–25] in a treatment time-dependent [23,26] manner. LTPs have thus been developed for various therapeutics [24,25]. The LTP components interacting with the DNA origami tiles in this case are reactive neutrals and photons (no charged particles or electric fields).

Our findings indicate that the physical properties of DNA origami carriers must be considered to ensure optimal drug loading and have implications for the design of DNA origami systems for delivery of intercalated drugs and dose control.

Materials and Methods

*1.1 Reagents*

Phosphate buffer (PB), pH7: 39 parts 0.2M monobasic sodium phosphate monohydrate and 61 parts 0.2M anhydrous dibasic sodium phosphate. Purification buffer: 10mM Tris-HCl, 1mM EDTA, 50mM NaCl, 10mM $MgCl_2$. Synthesis buffer: 1xTAE buffer (40 mM Tris-acetate and 1 mM EDTA, pH 8.3), 12.5mM magnesium acetate.

*1.2 DNA origami*

Single-stranded m13mp18ss scaffold DNA (New England BioLabs) was mixed with approximately 100 x excess of the non-biotinylated staple strands (Integrated DNA Technologies, Inc.) in synthesis buffer and annealed from 95ºC to room temperature, 20˚C, at a cooling rate of 1ºC per minute. After assembly, biotinylated strands were added at approximately 100 times the scaffold concentration and the mixture was hybridised overnight in synthesis buffer at room temperature.

The individual staples and their complements used in gel electrophoresis experiments were annealed by heating to 95˚C and cooling to room temperature at a rate of 1˚C per minute to form a 10 µM solution in synthesis buffer.

*1.3 Purifying Origami*

Excess staples were removed from the DNA origami using high resolution Sephacryl S-300 media (GE Healthcare) packed into filtration columns [27,28]; Full details of the preparation of the filtration media and packing of the spin columns are given in the supplementary methods.

*1.4 DNA origami surface immobilization for microscopy*

Coverslips (thickness 0.13-0.17mm, MNJ-350-020H, Menzel Gläser) were functionalised as follows for fluorescence microscopy of immobilised DNA origami. Coverslips were plasma cleaned (Harrick PDC-32G) for 5 minutes then soaked in BSA-biotin (0.5 mg/ml in PB) for 1 hour by pipetting 150-200 µl of liquid onto each coverslip. Each coverslip was rinsed twice with PB and air dried (after this stage in the preparation protocol dried coverslips can be stored at room temperature in Petri dishes sealed with Parafilm M (Bemis Company, Inc.) for later use as appropriate). At all times the cleaned/functionalised side of the coverslip was maintained upwards such that the non-functionalised side was in contact with the Petri dish or equivalent surface. Tunnel slides for microscopy were created by laying two parallel strips of Scotch double sided tape (product number: 70071395118, 3M) spaced 2-4 mm apart on a slide and placing a functionalized coverslip on top. The coverslip was tapped down with a pipette tip avoiding the area to be imaged and the excess tape removed. The tunnel was hydrated with 20 µl PB then incubated with 10 µl of 1 mg/ml avidin (pH7) for 1 hour. Excess avidin was removed by washing with 20 µl purification buffer (using a kimwipe and capillary action, without letting air bubbles through) before the addition of 7.5 µl of purified biotinylated origami in purification buffer (concentration approximately 0.8 nM), incubated for 5 minutes and washed with 20 µl of purification buffer. Fluorescent dye labelling was performed by flowing 10 µl of YOYO-1 diluted 1:199 in purification buffer and incubating for 5 minutes. Excess dye was removed by washing with 100 µl purification buffer. All incubation steps were performed at room temperature in a humidity chamber.

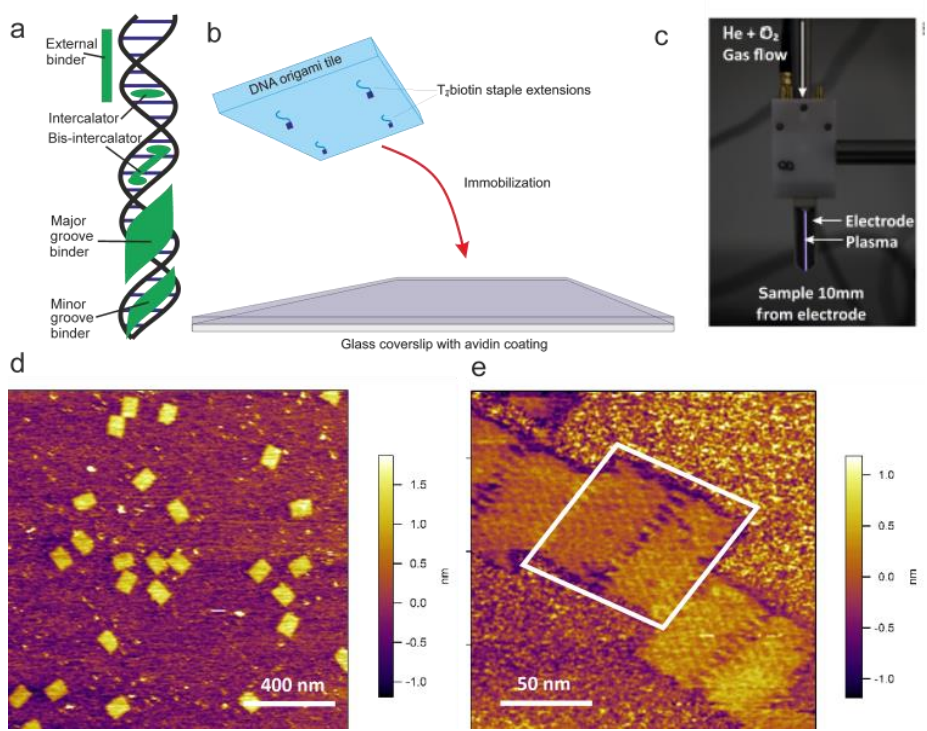

Figure 1: Experimental methods. (a) Schematic diagram of DNA binding modes. (b) Schematic diagram showing the method of DNA origami immobilization. (c) Image of the low temperature plasma source used to induce DNA damage (d,e) AM-AFM topography images of undamaged DNA origami tiles in liquid.

### 1.5 Low temperature plasma treatment of DNA origami

Samples of approximately 0.8 nM purified DNA origami (estimated based on pre-purification concentration) were treated with LTP for 1 minute. Samples of 20 µl were treated in the lids of 500 ml Eppendorf tubes, without the tube attached, and were then placed centrally [26] and vertically 10 mm underneath the electrodes. The low temperature plasma source was a prototype of the COST Reference Microplasma Jet [29] (1 mm electrode gap and width, 30 mm length) and operated using an applied 13.56 MHz peak-to-peak voltage of 450-500 V, with a feed gas of 1 slm (standard litre per minute) helium and a 0.5% admixture of oxygen.

### 1.6 Fluorescence microscopy

Fluorescence microscopy was performed on a bespoke optical imaging system, built around a commercial microscope body (Nikon Eclipse Ti-S) as previously described [30]. A supercontinuum laser (Fianium, SC-400-6, Fianium Ltd.) coupled to an acousto-optic tunable filter (AOTF) set to 45% at a central wavelength of 491 nm was used for illumination (see supplementary figure 1 and supplementary note), giving a power density at the sample of 775 Wcm$^{-2}$. A 475/50 nm wavelength bandpass filter was used to provide blue light excitation. The beam was de-expanded to generate narrowfield epifluorescence illumination [31] with FWHM of laser excitation at the sample of approximately 12 µm. A 515 nm wavelength dichroic mirror and 535/15 nm wavelength emission filter were used in the filter set. A 100x objective lens (Nikon oil imTIRF objective NA 1.49) was used with a further 2 times magnification to give a total magnification of 120 nm pixel$^{-1}$ at the emCCD camera detector used (Andor iXonEM+ DU 860 camera, Andor Technology Ltd).

Images were acquired at 10 ms exposure times (corresponding to a frame time of 10.06 ms), at pre-amplifier and EM gains of 4.6 and 300 respectively.

### 1.7 Gel electrophoresis

50ml 1% agarose (analytical grade, Promega) gels pre-stained with SYBR safe (Invitrogen) or post-stained with acridine orange were run horizontally at 100V (6.7Vcm$^{-1}$) in TAE buffer, for 30 minutes and imaged via an automated gel imaging system (ChemiDoc MP Imaging System, Bio-Rad Laboratories). DNA samples were prepared in 6 µl volumes with 0.6 µl of 10x running buffer, 0.75 µl filtered 80% glycerol (0.45 μm diameter pore syringe filter), and 1μl of 6X purple loading dye (for pre-stained gels) or 1 μl of ultrapure water (for post-stained gels). The rest of the sample volume comprised 50-100 ng DNA or DNA diluted in ultrapure water.

For SYBR safe pre-stained gels, 5 μl of 10,000x concentrated SYBR Safe was mixed via gentle swirling into the agarose before casting. For acridine orange post-staining gels were run in TAE buffer without stain. The post-staining protocol was adapted from McMaster and Carmichael [18]; gels were incubated at room temperature with rocking frequency 0.25 Hz for 30 minutes in 50 ml of 30 μg/ml acridine orange in TAE, to completely submerge the gel. The buffer was exchanged for 1 hour destain in 50 ml TAE under rocking incubation at 0.25 Hz.

Images were analysed to determine ratios of red to green fluorescence (RG ratio) and standard deviation using custom written Matlab routines (available at http://single-molecule-biophysics.org/ ). Each band to be analysed was made into a sub-image in both the red and green channels. Both images were morphologically opened using a disk of radius 2 pixels, thresholded using Otsu's method [32], morphologically opened with a disk of radius 1 pixel to remove small artefactual segmentation holes and then dilated to produce masks of the band areas. These masks were correlated to determine the overlapping area, which was used to segment each band (see supplementary figure 2). The band standard deviation was calculated by multiplying the number of pixels in the band by the per pixel standard deviation. In some images dust in the original image had to be removed before quantification: a 15 pixel horizontal line profile was taken through pixels seen to contain dust and fitted using a 5$^{th}$ order polynomial, with zero weight for the dust affected pixels. The residuals for the dust affected pixels were removed from their intensity values before calculation of the red to green fluorescence ratio. All pre- and post-correction images can be seen in supplementary figure 3.

*1.8 Particle Tracking and determination of single dye Intensity*

A bespoke single-molecule precise tracking and quantification software called ADEMS code [16] was used to evaluate the fluorescence intensity of the tiles over time. As DNA origami tiles are immobilised by biotin conjugation the detected spots were linked into fluorescent trajectories in the code based on their location in the kinetic series of image data. Images were corrected for non-uniform illumination using a profile of the beam created by raster-scanning 200 nm yellow-green fluorescent beads (F8811, ThermoFisher Scientific) through the focal volume. Objects within a 40 pixel radius of the beam centre were used to ensure a flat field of illumination, minimising the effect of noise. Objects containing more than 20 localizations in the first 100 image frame intensities were fitted using an exponential function of the form:

$$I(t) = I_0 e^{-Bt} + C$$

Where *C* is an offset to account for the presence of photoblinking behaviour of YOYO-1 at long times [16], B is the characteristic photobleaching time and $I_0$ is the height of the photobleaching exponential above the photoblinking behaviour. The initial intensity was taken as *C+I$_0$*; any saturated image frames were excluded from the analysis with the exponential function fitted to the remaining points.

The single-molecule YOYO-1 fluorescence intensity was determined using intensity *vs.* time traces for DNA origami tiles sparsely labelled with YOYO-1. Multiple analysis techniques were used to determine a consensus single-molecule intensity (see supplementary methods and supplementary figure 4): peak positions in the histogram of fluorescence intensity of origami over time; peaks in the pairwise intensity of the same data, fast Fourier transforms of the pairwise intensity, and Chung-Kennedy filtering [33,34] of individual traces with two window sizes to allow for the photobleaching and photoblinking effects of YOYO-1 were all used to determine the single YOYO-1 intensity of approximately 370 detector counts (range 356-380).

*1.9 Calculation of stoichiometries*

Kernel density estimates [35] of the initial intensity distribution of single tiles divided by the characteristic intensity of a YOYO-1 molecule were used to determine mean stoichiometries. The locations of peaks were determined objectively using the Matlab findpeaks.m function. The full width half maximum (FWHM) of the first peak in the kernel density estimate was fitted using a Gaussian distribution to generate the mean stoichiometry (i.e. number of detected dye molecules present within each detected fluorescent spot), with the sigma width used as an estimate for the error on the mean stoichiometry. In the case that a second peak was identified by the findpeaks.m function within the FWHM of the first peak the data was truncated to include only the first peak.

*1.10 Atomic force microscopy*

AFM imaging was performed in amplitude modulation (AM) mode in fluid with a MFP-3D (Oxford Instruments Asylum Research Co., Ltd, UK) using silicon nitride RC800 cantilever tips (Olympus) with nominal spring constant 0.4 $Nm^{-1}$. 5 μl of DNA origami tiles were incubated at room temperature in purification buffer on freshly cleaved mica for 15 minutes. Images were taken in purification buffer at room temperature at a scan speed of 1 Hz.

Before performing AFM of LTP damaged DNA origami tiles we verified that no further damage to DNA occurs after LTP treatment is removed (supplementary methods, supplementary figures 5,6).

**Results**

Fluorescence measurements of acridine orange bound to DNA origami indicate a non-intercalating binding mode

Acridine orange fluoresces red when bound to single stranded DNA via electrostatic binding to phosphates [21], and green when bound to double stranded DNA via an intercalative mode [19,20]. To investigate the accessibility of the DNA origami structure various single- and double-stranded DNA constructs and the DNA origami tiles were subjected to agarose gel electrophoresis and post-stained with acridine orange (figure 2 and supplementary figure 7).

M13mp18ss (the viral DNA plasmid used as the DNA origami tile backbone) is single stranded and appeared red in the merged fluorescence image (figure 2a), producing a red to green fluorescence intensity ratio (RG ratio, see methods section 1.7) of 21.54 ± 4.53 (± standard deviation; figure 2b) when imaged in red and green fluorescence using a gel imager (ChemiDoc MP Imaging System, Bio-Rad Laboratories). The double stranded pUC19 plasmid appeared green in the fluorescence images and gave an RG ratio of 0.33 ± 0.02. The undamaged DNA origami tile monomer showed intermediate behaviour, appearing neither red nor green in the merged fluorescence images and producing an RG ratio of 5.10 ± 0.83. A table of RG ratios for different DNA constructs is given in table 1 and shown in figure 2b.

| DNA sample | Double- (ds) or single- (ss) stranded | Red to green fluorescence intensity ratio (± SD) |
|---|---|---|
| m13mp18ss | ss | 21.54 ± 4.53 |
| 3kbp band; 2-log ladder | ds | 0.32 ± 0.01 |
| pUC19 | ds | 0.33 ± 0.02 |
| Origami Monomer | ds | 5.10 ± 0.83 |
| Origami Dimer | ds | 8.93 ± 2.02 |

Table 1: Red to green fluorescence intensity ratios of DNA constructs run in agarose gel electrophoresis and post-stained with acridine orange, as shown in figure 2d, with standard errors (SE).

As further controls, two single stranded DNA staples from Rothemund's original DNA origami tile (r3t10f and r3t22f; see table 2) were also run in the gel, along with their complements and duplexes made by annealing the staples and complements (figure 2c). According to predictions from the NUPACK structure prediction software [34], staple r3t10f has a low probability of forming secondary structure whilst r3t22f has a high probability of forming secondary structure. Therefore r3t10f was expected to be mostly single-stranded, while r3t22f was expected to be mostly double-stranded.

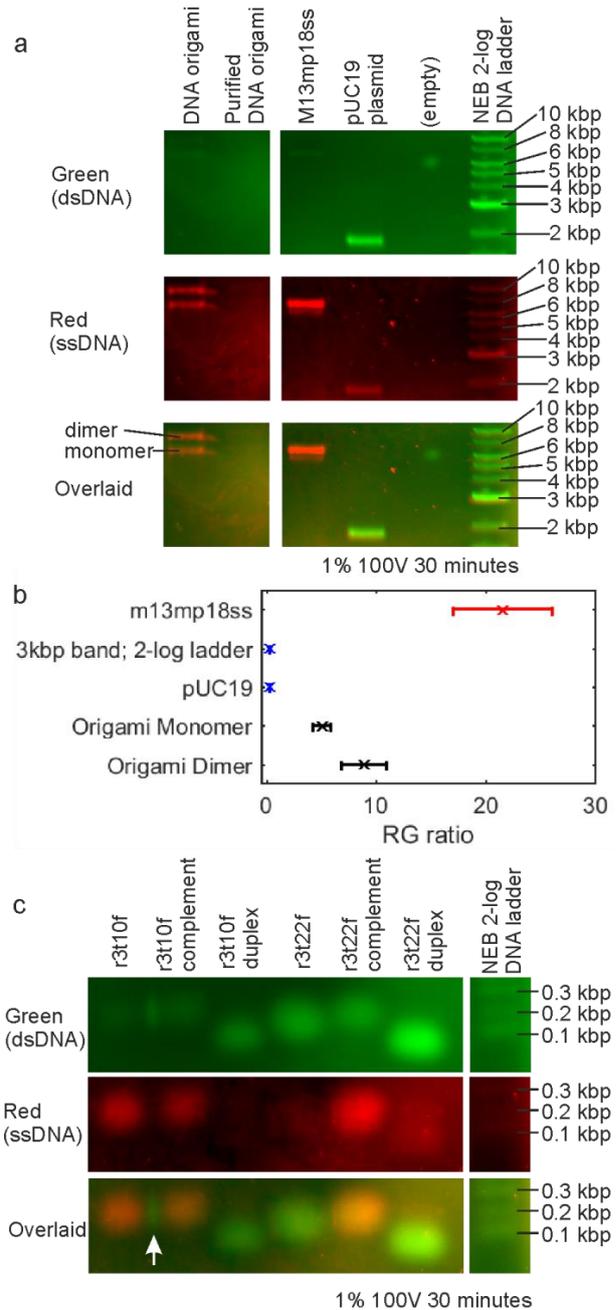

| Staple name | Sequence |
|---|---|
| r3t10f | 5'−ATTATTTAACCCAGCTACAAT TTTCAAGAACG−3' |
| r3t10f complement | 5'−CGTTCTTGAAAATTGTAGCTG GGTTAAATAAT−3' |
| r3t22f | 5'−AGGCGGTCATTAGTCTTTAAT GCGCAATATTA−3' |
| r3t22f complement | 5'−TAATATTGCGCATTAAAGACT AATGACCGCCT−3 |

Table 2: Oligonucleotides used in gel electrophoresis.

Staple r3t10f is expected to have single stranded behaviour and fluoresces red, as does its complement. The high mobility of these staples allows them to diffuse beyond their wells, and r3t10f and its complement are seen to hybridise in the overlapping region, indicated by the green fluorescent region between these two bands (figure 2c, white arrow). The annealed duplex of these bands fluoresces green indicating double stranded behaviour. As the non-annealed duplex occurs at the same position on the gel as the single strands it is likely that this was formed due to diffusion of the low molecular weight single stranded staples during staining, after the removal of the electrophoretic voltage.

Figure 2: Acridine orange interaction with DNA origami imaged via gel electrophoresis. (a) Green, red and merged fluorescence intensity images of DNA samples in electrophoretic gel. Purified DNA origami bands are faint. (b) Red to green fluorescence (RG) ratios with standard deviations of bands from the gel shown in part (a): single stranded species are shown in red, double stranded in blue and the intermediate origami results in black. (c) Green, red and merged fluorescence intensity images of DNA staple samples in electrophoretic gel. White arrow indicates hybridisation of r3t10f and r3t10f complement without annealing. All gel images are from the same gel shown in full in supplementary figure 7. All images in a colour channel are shown at the same contrast levels.

Staple r3t22f, expected to have high self-complementarity and to form secondary structure, fluoresces green, indicating that it is indeed double stranded, whilst its complement is single stranded and its duplex is double stranded as predicted. For this staple no hybridisation with its complement is seen in the region between the bands, consistent with the expectation that annealing is required to remove secondary structure and allow the two staple strands to hybridise.

The RG ratios for the individual staples run as controls and related complexes were not evaluated as their high mobility in the gel produced diffuse bands that extended beyond the original lanes and were of lower contrast than the higher molecular weight bands.

All control samples with known single- or double-stranded behaviour showed an RG ratio consistent with the predicted binding mode for this structure, whilst the DNA origami tile monomer and dimer showed an RG ratio intermediate between the intercalative and phosphate-binding modes.

The DNA origami tile is seen to be well-formed into a double helical structure in AFM images (see figure 1d, e). However, the intercalating DNA binding mode of acridine orange is not observed to be dominant upon post staining of an agarose gel. The RG ratio is between the values observed for ssDNA or dsDNA, which may indicate that both modes of binding are present. Other possible explanations are concentration-dependent binding effects of acridine orange [36] but this is unlikely for the concentration ranges used in this work and would be inconsistent with the observed results for controls of known strandedness. There is a region of the M13mp18ss scaffold which is left unstapled in the DNA origami synthesis, but this region is likely to contain large amounts of secondary structure [22] and be double stranded, so phosphate binding of acridine orange to this region is unlikely to explain this result. The t-loops added to prevent end-stacking result in the addition of 116 single bases per DNA origami tile, a small number compared to the expected 6480 base pairs. The t-loops may be responsible for the observed single-stranded binding mode, but cannot explain the lack of fluorescence from the double-stranded binding mode.

Gel electrophoresis indicates that acridine orange mostly binds predominantly in its single-stranded DNA binding mode to the double-stranded DNA origami tile despite the DNA origami tile being well formed; one possible explanation is that steric effects of DNA packing may reduce the accessibility of the DNA intercalation sites.

Single-molecule optical microscopy of the intercalating dye YOYO-1 labelled DNA indicates a saturating stoichiometry consistent with steric exclusion of intercalation

To test the accessibility of the DNA origami tile helices to intercalation we used the well-studied bis-intercalator YOYO-1. This dye has a reported maximum intercalative labelling of one YOYO-1 molecule to every four base pairs [13,37], corresponding to one moiety per two base pairs. A higher loading of one YOYO-1 every three base pairs [38] has been reported, consistent with recent studies using ethidium bromide, which challenge the idea that intercalating dyes completely exclude intercalation into neighbouring sites [39].

For the DNA origami tile used in this work, the number of intercalating YOYO-1 molecules per tile can be estimated using fluorescence microscopy in the saturating dye regime; the initial fluorescence intensity of a DNA origami tile can be divided by the total summed pixel fluorescence intensity (i.e. the observed brightness) of a single YOYO-1 molecule (see methods and supplementary methods) to estimate the number of dye molecules loaded.

The number of dye molecules required to reach the saturating regime of YOYO-1 intercalation was investigated using lambda DNA (supplementary figure 8, supplementary methods) and was found to saturate between 0.13 and 1.3 YOYO-1 molecules per base pair. Using these ratios as a guide, but considering the evidence for reduced intercalation into the DNA origami tile, a range of concentrations, 0.05-5μM, was tested for fluorescence intensity saturation on the undamaged DNA origami tile (figure 3, supplementary figure 9, table 3).

The saturating YOYO-1 concentration was found to be approximately 5μM; at this concentration the mean occupancy of intercalated YOYO-1 (± standard deviation) was found to be 67±25 molecules per DNA origami tile. The DNA origami tile contains approximately 6480 base pairs so we would expect a saturating concentration of roughly 1620 molecules, assuming maximum loading. Alternatively, if only the outward facing half of the outermost helices were accessible for dye binding we would expect a loading of ~68 molecules, given the reported maximum intercalative labelling of one YOYO-1 molecules per 4 base pairs.

The structure of this DNA origami tile is known to be tightly packed, with measured helix separations of 0.9-1.2 nm [22], whilst the size of a YOYO-1 analogue, TOTO, identical except for a sulphur atom in place of an oxygen, is larger than this separation, at around 2 nm [40], and YOYO-1 is known to require more room than TOTO in intercalation sites [41]. Whilst intercalation of YOYO-1 to DNA is known to perturb the structure of single DNA helices with each bis-intercalation unwinding the helix by 106º [41] and causes an extension of $0.68 \pm 0.04$ nm [42], it is not clear whether this effect or the precise positioning of DNA helix crossovers in this DNA origami tile would have more influence on the DNA origami structure: non- B form structure imposed in a DNA origami tile is known to change the binding properties of minor groove binding dyes [43].

The low intercalation of YOYO-1 into the DNA origami tile combined with the non-intercalative binding of acridine orange to the DNA origami tile suggests that the inner helices of this DNA origami tile are inaccessible to intercalation.

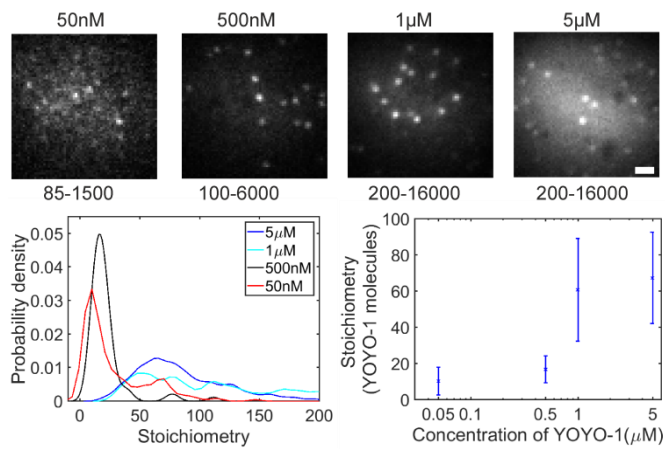

Figure 3: Single-molecule microscopy of the intercalating dye YOYO-1 with DNA origami. (a) Fluorescence microscopy images of DNA origami labelled with different concentrations of YOYO-1, before correction for illumination profile, with contrast levels used for display. These images are shown at the same contrast levels in supplementary figure 9. Scale bar 1 μm. (b) Kernel density estimate of the stoichiometry of YOYO-1 labelling on DNA origami tiles. (c) Peak stoichiometry (± standard deviation from Gaussian fit) as a function of YOYO-1 concentration from the data in (b).

| YOYO-1 concentration (μM) | Number of YOYO-1 molecules intercalated (±SD of Gaussian) | Number of DNA origami tiles |
|---|---|---|
| 0.05 | 10±8 | 169 |
| 0.5 | 17±7 | 67 |
| 1 | 61±28 | 326 |
| 5 | 67±25 | 452 |

Table 3: YOYO-1 binding to DNA origami in different concentrations, rounded to whole numbers. Number of YOYO-1 molecules and standard deviation (SD).

Inducing low temperature plasma damage to the DNA origami increases the number of YOYO-1 molecules per DNA origami tile observed by fluorescence microscopy, consistent with exposing more of the DNA structure.

If the inner DNA helices of this DNA origami tile are inaccessible to YOYO-1 intercalation due to steric hindrance, then reducing the structure to increase helix separation should increase dye intercalation.

To test this hypothesis low temperature plasma was used to induce single- and double-strand breaks in the DNA. An increase in the stoichiometry measured by single-molecule counting of YOYO-1 labelled DNA when DNA is damaged in this way would indicate that more structure becomes accessible after damage. If the entire structure is already accessible, no increase in stoichiometry would be seen.

Previous work with LTP has shown that small levels of DNA damage are typically inflicted over timescales of a few seconds to a few minutes [23,25]. The 60 s duration of treatment was determined to be a suitable LTP dose to inflict damage on the DNA origami tiles whilst maintaining some structure (supplementary methods, supplementary figures 10,11).

DNA origami tiles which had been treated with low temperature plasma for 60 s and labelled with YOYO-1 showed a higher intensity than untreated DNA origami tiles in a fluorescence microscopy assay (see figure 4a, b, table 4). This observation indicates that more of the DNA origami structure is likely to be accessible to intercalation of YOYO-1.

AFM images of the 60 s LTP treated DNA origami tiles (figure 4c), reveal that many of the damaged DNA origami tiles are fractured or incomplete. Few holes in the DNA origami are seen, implying that single staples are not removed during damage, as omission of single staples has previously been shown to produce small holes in the DNA tiles [22].

| DNA Origami tile state | Number of YOYO-1 molecules intercalated | Number of DNA origami tiles |
|---|---|---|
| Undamaged | 67±25 | 452 |
| 60 s LTP treatment | 108±12 | 224 |

Table 4: Number of YOYO-1 molecules bound to DNA origami tiles at a 5 μM YOYO-1 concentration in different LTP treatment states, rounded to whole numbers and standard deviation.

Together, these fluorescence microscopy and AFM results indicate that LTP does break the DNA origami tiles and creates more YOYO-1 accessible base pairs within a single tile, supporting the idea that intercalation is strongly affected by steric interactions.

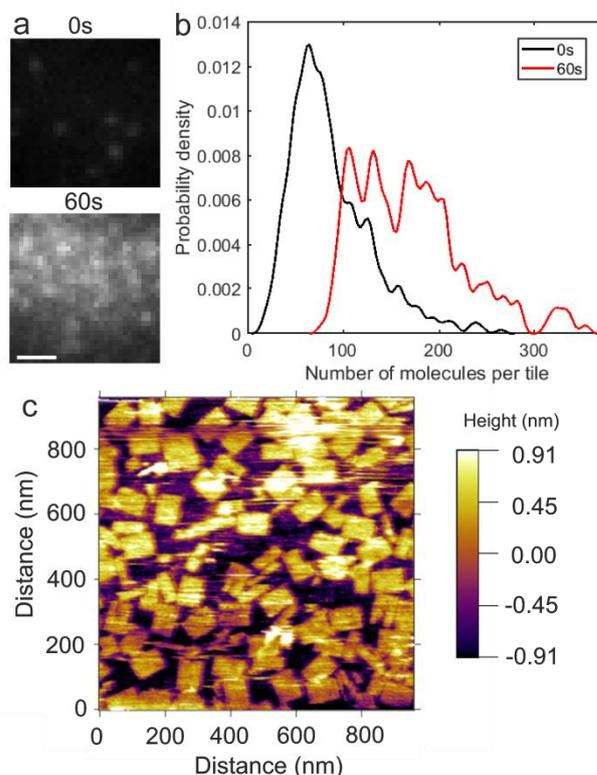

Figure 4: LTP damages the DNA origami tiles allowing greater YOYO-1 binding. (a) Representative fluorescence images of YOYO-1 binding to undamaged and 60s LTP treated DNA origami tiles. Scale bar 1 μm (b) Kernel density estimate of fluorescence intensity of YOYO-1 emission of DNA origami tiles before and after LTP treatment. (c) AFM images of 60s LTP treated DNA origami tiles.

**Conclusions**

Taken together, our results show that the inner helices of the DNA origami tile used in this work are likely to be inaccesible to intercalating binding, but that higher loading can be achieved when the tiles are damaged to expose more DNA helices; we suggest that intercalation is prevented in the undamaged tile by either steric or electrostatic effects due to the proximity of the helices, but this may depend on intercalator size, which was not investigated here. DNA origami nanostructures are promising candidates for use as drug-delivery vehicles for mediating targeted delivery due to their high biocompatibility; factors affecting drug loading, such as those outlined here, have important ramifications for dose control. Recent work has shown that large DNA origami tiles with a high external surface area-to-volume ratio are the most easily taken up into cells, but our results suggest that these DNA origami structures may be sub-optimal for drug delivery due to unavailability of drug binding sites, and open, mesh-like structures [44,45] might be better suited for intercalating drug delivery due to a higher proportion of the bases being intercalator accessible.

## Acknowledgements

We are grateful for discussions with Dr Steve Johnson (Department of Electronics, University of York) concerning surface chemistry immobilisation methods, and for use of resources in the Bio-Inspired Technology Laboratory. MCL was supported by the BBSRC (grants BB/N006453/1, BB/P000746/1 and BB/R001235/1) and the EPSRC (grants EP/T002166/1). KD's work on this paper was part-funded by the EPSRC Platform Grant EP/K040820/1.